# Outlining where Humans live – the World Settlement Footprint 2015

**Mattia Marconcini**[1*], **Annekatrin Metz-Marconcini**[1], **Soner Üreyen**[1], **Daniela Palacios-Lopez**[1], **Wiebke Hanke**[1], **Felix Bachofer**[1], **Julian Zeidler**[1], **Thomas Esch**[1], **Noel Gorelick**[2], **Ashwin Kakarla**[3], **Emanuele Strano**[4]

[1] German Aerospace Center (DLR), Wessling, Germany
[2] Google Switzerland, Zurich, Switzerland
[3] Google India, Hyderabad, India
[4] MindEarth, Biel/Bienne, Switzerland

October 25, 2019

## Abstract

Human settlements are the cause and consequence of most environmental and societal changes on Earth; however, their location and extent is still under debate. We provide here a new 10m resolution (0.32 arc sec) global map of human settlements on Earth for the year 2015, namely the World Settlement Footprint 2015 (WSF2015). The raster dataset has been generated by means of an advanced classification system which, for the first time, jointly exploits open-and-free optical and radar satellite imagery. The WSF2015 has been validated against 900,000 samples labelled by crowdsourcing photointerpretation of very high resolution Google Earth imagery and outperforms all other similar existing layers; in particular, it considerably improves the detection of very small settlements in rural regions and better outlines scattered suburban areas. The dataset can be used at any scale of observation in support to all applications requiring detailed and accurate information on human presence (e.g., socioeconomic development, population distribution, risks assessment, etc.).

***Keywords*** Global Urbanization, Settlement Extent Mapping, Earth Observation, Sentinel-1, Landsat-8

## Background & Summary

Scientific investigations related to the human presence on Earth strongly rely on the availability of accurate and reliable information on the extent and location of settlements. In this framework, since early 1980s satellite imagery has been used as primary source to outline settlements at global scale[1,2] and - along with technical, methodological and computational advances - their spatial detail evolved from low resolution (1km - 500m) to medium resolution (100m) and, since the last few years, to high resolution (30-10m).

[*] Corresponding Author: mattia.marconcini@dlr.de

Nowadays, satellite-based settlement extent maps are widely used for many scientific purposes. For instance, in global urbanization analyses[3] they are used to define urban areas, as well as characterize their morphology and assess the correlation with socio-economic variables. In spatial demography[4], the location of settlements represents a fundamental spatial covariate to model the displacement of people, whereas in land-use science[5,6] settlement extent is exploited as key input for calibrating land-use change models.

At present, the most largely employed layers outlining the global settlement extent include:

- the Global Urban Footprint – GUF[7] (available at 12m resolution and referring to the year 2012) generated by the German Aerospace Center (DLR) from 3m resolution TerraSAR-X/TanDEM-X radar imagery;
- the 2014 instance of the Global Human Settlement Layer – GHSL[8], generated at 30m resolution by the Joint Research Center (JRC) of the European Commission from Landsat-8 optical imagery;
- the artificial surfaces mask of the GLOBELAND30 – GLC30[9] (available at 30m resolution and referring to the year 2010), generated by the National Geomatics Center of China from Landsat-5/7 optical imagery.

Among these, the GUF outperforms the other two layers[10], which show severe under- and over-estimation in large parts of the world. Nevertheless, the GUF itself still exhibits two major drawbacks. On the one hand, it has been generated (like the GHSL and the GLC30) from single-date scenes, which are sometimes strongly affected by the specific acquisition conditions, hence resulting in misclassification errors. On the other hand, commercial imagery has been employed, which prevents a systematic update due to its high costs. Moreover, the exclusive use of optical or radar imagery alone represents an additional limitation since these two types of data are sensitive to different structures on the ground (i.e., artificial surfaces and built-up areas, respectively). For instance, bare soil and sand tend to be misclassified as settlements using optical imagery, while this does generally not occur with radar data; on the contrary, complex topography areas or forested regions can be wrongly categorized as settlements with radar imagery, whereas normally this does not occur using optical data.

To overcome these issues, we have developed a novel and robust methodology to reliably outline settlements which jointly exploits, for the first time, open-and-free multitemporal optical and radar data. In particular, the rationale is that the temporal dynamics of human settlements over time are different than those of all other non-settlement classes. First, we gather all the images acquired over a region of interest within a target period during which we do not expect considerable changes (e.g., one year). Next, we extract key temporal statistics (i.e., temporal mean, minimum, maximum, etc.) of: i) the original backscattering value in the case of radar data; and ii) different spectral indices (e.g., vegetation index, built-up index, etc.) derived after performing cloud/cloud-shadow masking in the case of optical imagery. After automatically extracting candidate training samples for the settlement and non-settlement class, binary classification based on advanced machine learning is separately applied to the optical- and radar-based temporal features. Finally, the two outputs are properly combined together.

Once tested its high robustness on a variety of study sites, the method has been employed to generate the World Settlement Footprint (WSF) 2015, a 10m resolution (0.32 arc sec) binary mask outlining the extent of human settlements globally derived by means of 2014-2015 multitemporal Sentinel-1 (S1) radar and Landsat-8 optical imagery (of which ~107,000 and ~217,000 scenes have been processed, respectively). The WSF2015 is extremely accurate and reliable and outclasses all other mostly employed similar datasets. This has been quantitatively assessed through an unprecedented validation exercise based on 900,000 ground-truth samples collected by crowdsourcing photointerpretation and carried out in collaboration with Google. To this purpose a statistically robust and transparent protocol has been defined following recommended state-of-the-art practices.

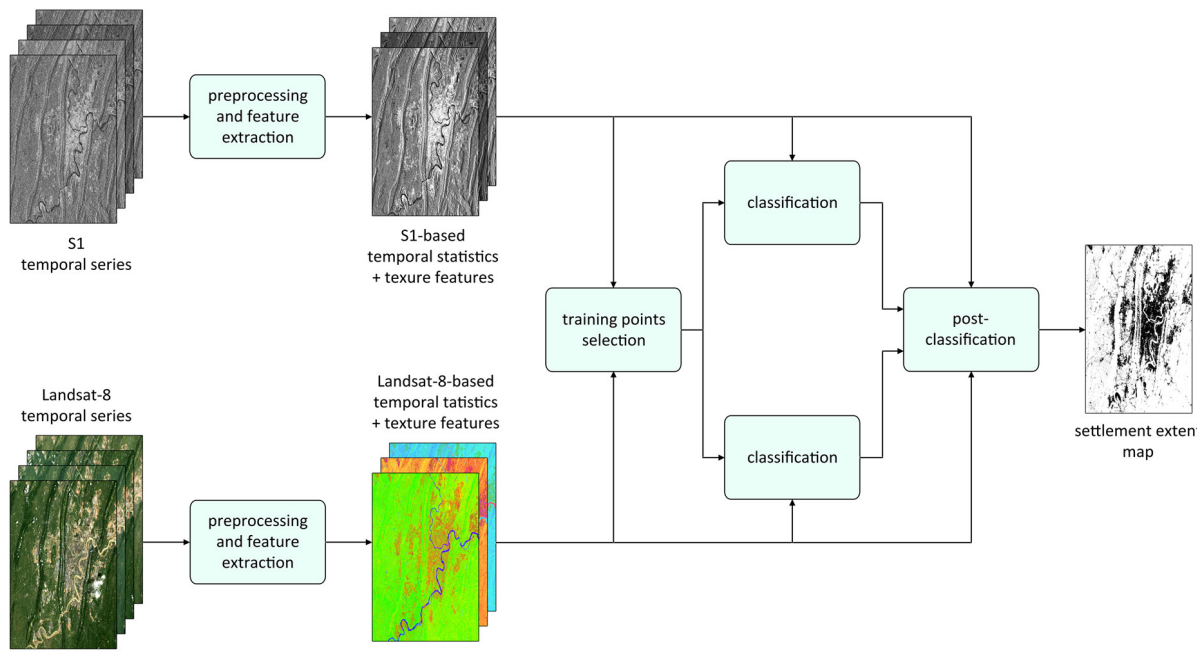

Figure 1. **Block scheme.** Schematization of the workflow implemented for outlining human settlement extent from Sentinel-1 (S1) radar and Landsat-8 optical multitemporal satellite imagery.

## METHODS

In this Section, we describe the novel methodology developed for outlining human settlement extent based on the joint use of multitemporal radar and optical imagery. The corresponding block scheme is reported in Figure 1. First, both S1 and Landsat-8 data are pre-processed and suitable temporal statistics and texture features are computed. Then, training points for the settlement and non-settlement classes are derived by jointly exploiting both radar- and optical-based temporal statistics (along with additional ancillary information). Classification is performed separately for the two types of data by means of an ensemble of Support Vector Machines (SVM) classifiers. A final post-classification phase is dedicated to properly combine the Landsat- and S1-based classification maps and automatically identifying and deleting potential false alarms.

Each of the abovementioned steps is described into detail in the following. Next, the WSF2015 layer is presented along with all relevant details concerning its implementation.

### Preprocessing and Feature Extraction

As concerns S1 data, we take into account imagery acquired in Interferometric Wide swath (IW) mode (i.e., S1 main mode over land with 250km swath). In particular, we consider High-Resolution Level-1 Ground Range Detected (GRD) products available at 10m resolution.

All scenes acquired over the given study area in the target timeframe are first gathered and then pre-processed by means of the S1 Toolbox[11]. Specifically, this task includes:

- orbit correction (for improving the geocoding);
- thermal noise removal (for removing dark strips near scene edges with invalid data);
- radiometric calibration (for computing backscattering intensity using sensor calibration parameters in the GRD metadata);
- Range-Doppler terrain correction (for removing the brightness and geometric distortions occurring in correspondence of elevated and sloping terrain);
- conversion to decibel (dB) values (for reducing the very high dynamic range of data).

| Spectral index | Formula |
|---|---|
| Normalized Difference Built-Up Index (NDBI) | (SWIR1-NIR)/(SWIR1+NIR) |
| Modified Normalized Difference Water Index (MNDWI) | (Green-NIR)/(Green+NIR) |
| Normalized Difference Vegetation Index (NDVI) | (NIR-Red)/(NIR+Red) |
| Normalized Difference Middle Infrared (NDMIR) | (SWIR1-SWIR2)/(SWIR1+SWIR2) |
| Normalized Difference Red Blue (NDRB) | (Red-Blue)/(Red+Blue) |
| Normalized Difference Green Blue (NDGB) | (Green-Blue)/(Green+Blue) |

Table 1. **Landsat-8 spectral indices.** Spectral indices extracted from Landsat-8 OLI imagery [Blue = band 2; Green = band 3; Red = band 4; Near Infrared (NIR) = band 5; Short-wave Infrared (SWIR) 1 = band 6; Short-wave Infrared (SWIR) 2 = band 7].

Scenes acquired in ascending and descending pass are treated separately due to the strong influence of the viewing angle in the backscattering of built-up areas. Furthermore, experimental analyses assessed that the joint employment of VV/VH imagery does not provide any considerable improvement with respect to the solely use of VV data; accordingly, VH data are disregarded.

As pointed out above, the rationale of the proposed approach is that given a series of multi-temporal images for a study area, the corresponding temporal dynamics of human settlements are sensibly different than those of all other non-settlement classes. For instance, in the case of radar data the backscattering temporal mean of built-up areas (due to double bounce reflection) is higher than that of forest areas (which might result in high backscattering in one/few acquisitions due to specific conditions, but in general exhibit lower values). To properly characterize this behavior, for each pixel we compute 5 key temporal statistics, namely the backscattering temporal maximum, minimum, mean, standard deviation, and mean slope (i.e., defined as the average absolute difference between consecutive items of the temporal series).

Texture information is also extracted to ease the identification of lower-density residential areas mostly characterized by single houses surrounded by vegetation (which are generally challenging to detect due to their lower backscattering values with respect to that of denser urban areas). To this purpose, we compute the coefficient of variation (COV) of the temporal mean backscattering, which is defined for each pixel as the ratio between the local standard deviation and the local mean calculated over a *NxN* pixel spatial neighborhood. In particular, the COV represents an estimate of the local image heterogeneity. Here, in the light of the 10m spatial resolution of the considered S1 data, a neighborhood of 5x5 pixels proved to be an effective choice.

Overall, both for VV ascending/descending passes, the final S1 feature stack includes 7 features, namely: the 5 abovementioned temporal statistics and the COV derived from the backscattering temporal mean, plus the number of available scenes per pixel.

In the case of Landsat-8, imagery taken at 30m resolution by the Operational Land Imager (OLI) sensor is used. In particular, we only consider scenes acquired in the target period over the study area with cloud cover lower than 60% (as reported in the corresponding metadata). Indeed, we experienced that further raising this threshold often results in accounting for images with non-negligible misregistration error. Data are then calibrated and Top-Of-Atmosphere (TOA) radiance is extracted.

A mask is then generated for each image to exclude pixels affected by cloud and cloud shadows from the analysis. To this purpose the Function of mask (FMask) algorithm is applied given its assessed effectiveness in the scientific community[12]. Besides pixels covered by clouds and cloud shadows, the algorithm also identifies snow, clear land and clear water pixels; in particular, this is done by jointly analyzing the Normalized Difference Vegetation Index (NDVI), the Normalized Difference Snow Index (NDSI), and the Brightness Temperature for the given scene.

A thorough experimental analysis has been carried out to identify a set of spectral indices highly suitable for an effective delineation of human settlements; in particular, the final list and corresponding formulas are reported in Table 1. The Normalized Difference Built-Up Index (NDBI)[13] has been applied to extract built-up areas in many studies[14,15]; nevertheless, due to the use of the first short-wave infrared (SWIR) band (i.e., OLI band 6) this index is also sensitive to vegetation with low water content[16], which exhibits values comparable to those of settlement areas. Accordingly, the Normalized Difference Middle Infrared index (NDMIR) and the NDVI are applied to overcome this issue. On the one hand, the NDMIR is computed using both SWIR bands (i.e., OLI bands 6 and 7), thus being sensitive to vegetation moisture[17]. On the other hand, the NDVI[18] has been widely employed in a variety of land cover applications as well as in the context of settlement extent classification[19,20]. Moreover, the Modified Normalized Difference Water Index (MNDWI)[21] is also employed to discriminate water from settlement areas. Such index enhances the performance of the NDWI[22] by replacing the MIR with the NIR band (i.e., OLI band 5), which leads to a reduction of noise from built-up areas. In addition to the previous, two other spectral indices have been included for improving the discrimination between settlement areas and bare soil/bare rocks; specifically, these are the Normalized Difference Red Blue (NDRB) and Normalized Difference Green Blue (NDGB) indices[23].

To characterize the generally stable temporal dynamics of the settlement class with respect to the other non-settlement classes, the same set of 5 key temporal statistics used in the case of S1 data are extracted for each of the 6 Landsat-8 spectral indices presented above. Moreover, to improve the detection of rural and suburban areas (mostly characterized by a low share of built-up areas and a high share of vegetation, thus resulting in a heterogeneous environment compared to denser built-up areas), also here additional texture features are extracted. In particular, for each of the derived 6 temporal mean indices, we computed the corresponding COV in a neighborhood of 3x3 pixels, which empirically proved the most effective choice in the light of the 30m spatial resolution of Landsat data.

The final Landsat-8 feature stack includes 37 bands, namely: temporal maximum, minimum, mean (plus the corresponding 3x3 COV), standard deviation and mean slope for NDBI, NDVI, MNDWI, NDMRI, NDRG and NDGB, along with the number of available cloud/cloud-shadow-free acquisitions per pixel.

In Figure 2, examples are given for Ho Chi Minh (Vietnam), Istanbul (Turkey), Johannesburg-Pretoria (South Africa), Karachi (Pakistan), Lagos (Nigeria), and Moscow (Russia). Specifically, in addition to reference Google Earth imagery we display for each of city: i) a RGB color composition obtained combining the Landsat-8 temporal mean NDBI (red), NDVI (green) and MNDWI (blue); ii) the S1 VV backscattering temporal mean. The same visualization parameters have been consistently applied to all 6 sites.

Yet by simple visual inspection, it is possible to appreciate the advantage of jointly employing radar- and optical-based temporal statistics. In particular, even if it is not feasible to properly delineate settlements by solely using radar data, it is clear how optical imagery helps overcoming this issue and vice-versa. For instance, in the case of Lagos, the backscattering is counterintuitively low in several highly urbanized areas; nevertheless, this occurs due to the extremely high building density (mostly informal housing) which prevents the typical radar double bounce reflection. Instead, when employing Landsat imagery the settlement outline is clearly distinguishable. On the contrary, optical-based temporal features are often not effective alone in arid regions - as in Karachi - where bare areas tend to be misclassified as settlements. Nevertheless, these can be effectively outlined by means of S1 temporal statistics.

**Training Points Selection**

Reliably identifying training points for the settlement and non-settlement class proved being the most critical task of the whole classification system; indeed, a training set including a consistent number of mislabeled samples would most likely result in poor performances. To this purpose, we designed a strategy which jointly exploits the temporal statistics computed for both S1 and Landsat data, along with additional ancillary information. In particular, any given sample $x$ in the study area is labelled as potentially settlement or non-settlement if it satisfies all the corresponding conditions listed in Table 2.

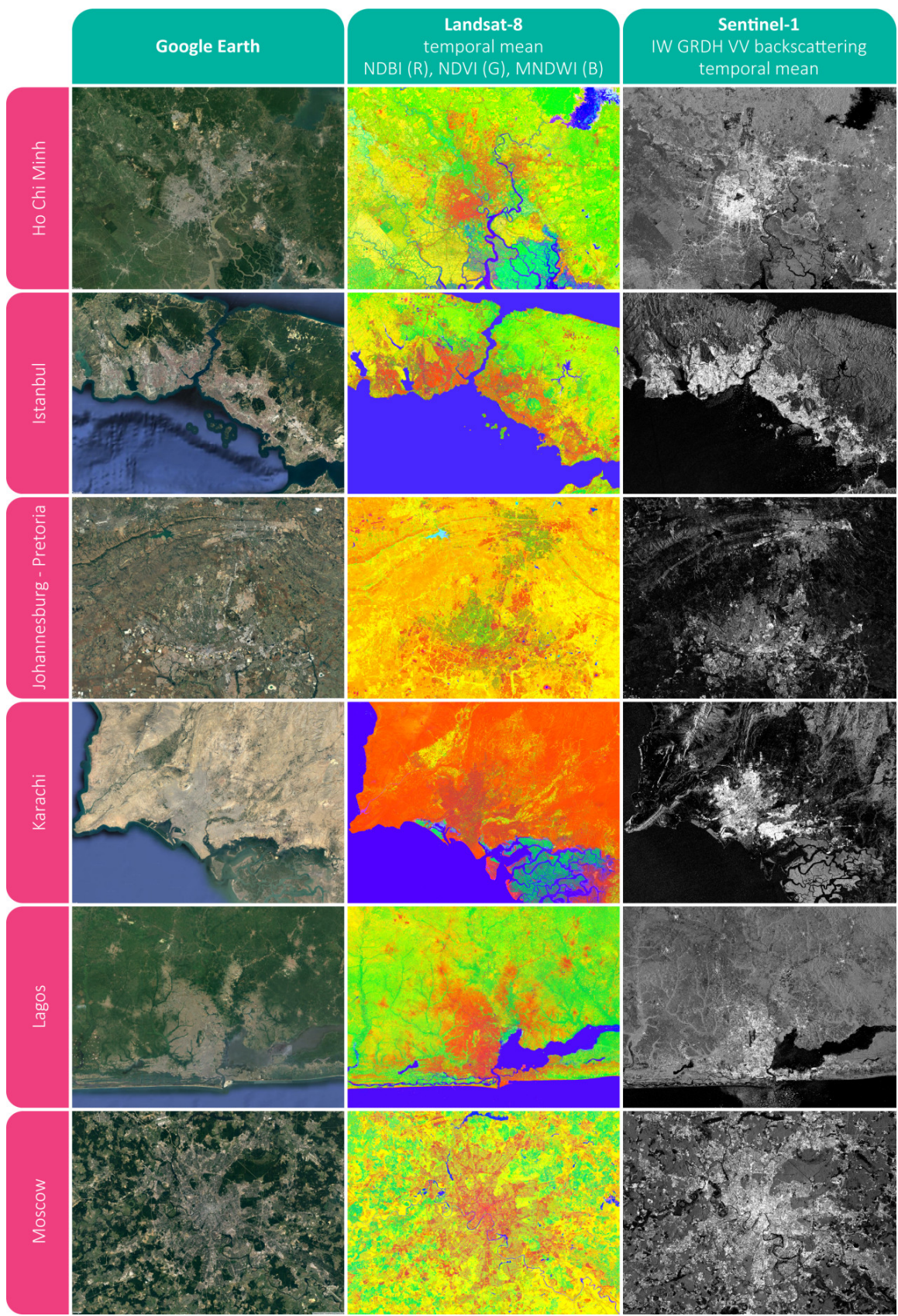

Figure 2. **Temporal features**. Examples for the cities of Ho Chi Minh (Vietnam), Istanbul (Turkey), Johannesburg-Pretoria (South Africa), Karachi (Pakistan), Lagos (Nigeria) and Moscow (Russia) including: i) Google Earth reference imagery; ii) RGB combination of the Landsat-8 temporal mean NDBI (Red), NDVI (Green) and MNDWI (Blue); and iii) Sentinel-1 IW GRDH VV temporal mean backscattering.

|  | **Candidate settlement pixels** | **Candidate non-settlement pixels** |
|---|---|---|
| Landsat-8 | $\begin{cases} \overline{NDBI}(x) > NDBI_{Smin}(KG(x)) \\ \overline{NDBI}(x) < NDBI_{Smax}(KG(x)) \end{cases}$ | $\begin{cases} \overline{NDBI}(x) < NDBI_{NSmin}(KG(x)) \\ \overline{NDBI}(x) > NDBI_{NSmax}(KG(x)) \end{cases}$ |
| Landsat-8 | $\begin{cases} \overline{NDVI}(x) > NDVI_{Smin}(KG(x)) \\ \overline{NDVI}(x) < NDVI_{Smax}(KG(x)) \end{cases}$ | $\begin{cases} \overline{NDVI}(x) < NDVI_{NSmin}(KG(x)) \\ \overline{NDVI}(x) > NDVI_{NSmax}(KG(x)) \end{cases}$ |
| Landsat-8 | $\begin{cases} \overline{MNDWI}(x) > MNDWI_{Smin}(KG(x)) \\ \overline{MNDWI}(x) < MNDWI_{Smax}(KG(x)) \end{cases}$ | $\begin{cases} \overline{MNDWI}(x) < MNDWI_{NSmin}(KG(x)) \\ \overline{MNDWI}(x) > MNDWI_{NSmax}(KG(x)) \end{cases}$ |
| Landsat-8 | $N_{LC8}(x) > 5$ | $N_{LC8}(x) > 5$ |
| S1 | $N_{S1A}(x) < 5 \vee \begin{cases} \bar{\sigma}_A^0(x) > -7\ dB \\ N_{S1A}(x) \geq 5 \end{cases}$ | $N_{S1A}(x) < 5 \vee \begin{cases} \bar{\sigma}_A^0(x) < -11\ dB \\ N_{S1A}(x) \geq 5 \end{cases}$ |
| S1 | $N_{S1D}(x) < 5 \vee \begin{cases} \bar{\sigma}_{tmD}^0(x) > -7\ dB \\ N_{S1D}(x) \geq 5 \end{cases}$ | $N_{S1D}(x) < 5 \vee \begin{cases} \bar{\sigma}_D^0(x) < -11\ dB \\ N_{S1D}(x) \geq 5 \end{cases}$ |
| DEM | $Slope < 10°$ | $Slope < 10°$ |

Table 2. **Training sample definition**. Criteria applied for outlining candidate settlement and non-settlement training samples.

Concerning optical data, we generally observed that most of the pixels can be effectively outlined as settlement/non-settlement by jointly thresholding the corresponding NDBI, NDVI, and MNDWI temporal mean features. Nevertheless, being all 3 spectral indices correlated to the presence of vegetation, absolute threshold values are not globally effective as vegetation strongly varies depending on climate. To overcome this drawback, we took into account the well-established Köppen Geiger scheme[24] and, based on extensive experimental analyses, for each climate type we empirically determined specific thresholds for outlining both candidate settlement and non-settlement training samples. Referring to Table 2, $KG(x)$ denotes the Köppen-Geiger classification for the given pixel $x$. $A_{Smin}(KG(x))$ and $A_{Smax}(KG(x))$ denote minimum and maximum thresholds, respectively, for the temporal mean $\bar{A}(x)$ of the spectral index $A, A \in \{NDBI, NDVI, MNDWI\}$ defined to determine whether $x$ is a candidate settlement training sample. Similarly, $A_{NSmin}(KG(x))$ and $A_{NSmax}(KG(x))$ denote minimum and maximum thresholds defined to determine whether $x$ is a candidate non-settlement training sample. Furthermore – in the reasonable hypothesis that the higher is the number of cloud/cloud-shadow free acquisitions, the more robust are the corresponding temporal statistics – we exclude all pixels whose number of Landsat-8 clear observations (i.e., $N_{LC8}(x)$) is lower than 5. Since the Köppen Geiger classification includes 30 different climate types, overall we determined 360 thresholds on the three indices.

Regarding radar data, we expect the temporal mean backscattering of most settlement samples to be sensibly higher than that of all other land-cover classes. Accordingly, samples whose temporal mean backscattering (either in the case of data acquired in ascending $\bar{\sigma}_A^0(x)$ and descending $\bar{\sigma}_D^0(x)$ pass) is:

- lower than -7 dB are not eligible to be labelled as settlement training samples (if the number of ascending/descending scenes used for computing the temporal statistics $N_{S1A}(x)$ / $N_{S1D}(x)$ is higher than or equal to 5);
- greater than -11 dB are not eligible to be labelled as non-settlement training samples (if the

number of ascending/descending scenes used for computing the temporal statistics $N_{S1A}(x)$ / $N_{S1D}(x)$ is higher than or equal to 5).

It is worth noting that in complex topography regions: i) radar data often show high backscattering comparable to that of settlements; and ii) bare rocks are present, which often exhibit a behavior similar to that of built-up areas in the Landsat-based temporal statistics. Accordingly, to exclude these from the analysis, we mask all pixels whose slope[25] (i.e., the angle corresponding to the maximum elevation difference between the given pixel and its 8 neighbors) is higher than 10 degrees. To this purpose, we employed the Shuttle Radar Topography Mission (SRTM)[26] Digital Elevation Model (DEM) for latitudes between -60° and +60° and the Advanced Spaceborne Thermal Emission and Reflection Radiometer (ASTER)[27] DEM elsewhere.

**Classification**

In the light of their proven effectiveness and high generalization capabilities, Support Vector Machines (SVM)[28,29] with Radial Basis Function (RBF) Gaussian Kernel have been chosen for the classification task.

In general, the criteria defined in the previous section result in a high number of candidate training points; thus, a subset should be sampled to keep the computational burden under control. For instance, a reasonable choice when investigating large regions is to subdivide the study area in working units of 1x1 degree size; in this case, an effective strategy proved extracting 500 samples for the settlement and 500 for the non-settlement class.

The stacks of Landsat- and S1-based temporal features are classified separately since this proved more effective than performing a single classification on the merger of the two stacks. In both cases, a grid search with a 5-fold cross validation approach is employed to identify the optimal values for the learning parameters (i.e., the ones expected to provide the best possible discrimination between the settlement and non-settlement classes). These include $\gamma$ and $C$ which tune the SVM kernel spread and error penalization, respectively[29]. In our analyses, we test all combinations with $C = 2^i, i \in \mathbb{Z}^{\geq 0}, i \leq 13$ and $\gamma = 0.1 \cdot j, j \in \mathbb{Z}^+, j \leq 20$.

Since results might vary depending on the specific subset of selected training points, as a means to further improve the final performances and obtain more robust classification maps, we randomly subset 20 different training sets and feed an ensemble of as many SVM classifiers. Then, we apply a majority voting approach[30,31] to handle the resulting maps and each pixel is finally associated with the settlement class only if it is labeled as settlement at least 11 over 20 times.

**Post-Classification**

A final post-classification phase is dedicated to properly combine the Landsat- and S1-based classification maps and automatically identifying and deleting false alarms. To this purpose, an updated version of the post-editing object-based approach adopted in the production of the GUF layer has been used[7], which exploits the 9 reference binary datasets (7 global and 2 continental) described in Table 3.

A settlement agreement mask is first generated from the combination of 6 reference layers (i.e., DLR-RC, CIL, OSM-S, OSM-R, GL30-S, and NLCD), which is labeled as positive only where two or more of these are positive. Likewise, a settlement exclusion mask is obtained by combining 3 reference layers (i.e., DLR-RM, GLC30-W, GLC30-WL), which is labelled as positive where at least one of these is positive.

Next, segmentation is applied to both Landsat- and S1-based classification maps for categorizing each cluster of connected pixels as individual objects; in particular, this is carried out by exploiting contour tracing to iterate over an image only once[32].

Objects are then removed if:

| Reference Layer | Description | Coverage |
|---|---|---|
| Relief Mask [DLR-RM] | Binary mask generated using the SRTM DEM for latitudes between -60° and +60° and the ASTER DEM elsewhere. It is labelled as positive where the shaded relief is greater than 212 or the roughness is greater than 15. | Global |
| OSM-Settlements [OSM-S] | Binary mask labelled as positive in correspondence of settlement-related OpenStreetMap geometries. | Global |
| OSM-Roads [OSM-R] | Binary mask labelled as positive in correspondence of road-related OpenStreetMap geometries. | Global |
| DLR Road Cluster [DLR-RC] | Binary mask obtained applying focal mean filtering to the OSM-R dataset. | Global |
| GLC30-Settlements [GLC30-S] | Binary mask labelled as positive in correspondence of GLC30 class 80 (i.e., artificial surfaces). | Global |
| GLC30-Water [GLC30-W] | Binary mask labelled as positive in correspondence of GLC30 class 50 (i.e., water). | Global |
| GLC30-Wetlands [GLC30-WL] | Binary mask labelled as positive in correspondence of GLC30 class 60 (i.e., wetlands). | Global |
| Copernicus Imperviousness Layer 2012 [CIL] | Binary mask labelled as positive where the Copernicus Imperviousness Layer 2012 exhibits values greater than 30%. | Europe |
| US National Land Cover Dataset 2011 [NLCD] | Binary mask labelled as positive in correspondence of classes 22, 23 or 24 from category "Developed" of the US National Land Cover Dataset 2011. | USA |

Table 3. **Reference layers**. Reference layers used in the post-classification phase.

- their extent overlaps for less than 30% the settlement agreement mask and, concurrently, it overlaps for more than 30% the settlement exclusion mask (this helps excluding objects wrongly covering complex topography regions, water or wetlands);
- the zonal mean of the Landsat-based temporal mean NDVI is higher than 0.6 (this is mostly the case of false detections in the S1-based classification occurring in correspondence of specific types of dense forests);
- the zonal mean of the S1 temporal mean backscattering (either computed for scenes acquired with ascending or descending pass) is lower than -11 dB (this is mostly the case of false detections in the Landsat-based classification occurring in correspondence of bare soil and sand).

The final classification map is given by the merger of the objects preserved in the Landsat- and S1-based classification maps.

**The WSF2015**

The methodology presented above has been applied globally to generate the WSF2015 layer. Concerning radar data, pre-processing and feature extraction have been performed for ~107,000 S1 scenes (i.e., ~51,000 collected with ascending pass and ~56,000 with descending pass ) acquired in 2014-2015. In particular, this task has been directly supported by Google through its Earth Engine cloud computing platform[33].

As regards optical imagery, pre-processing and feature extraction have been performed for ~217,000 Landsat-8 scenes acquired in 2014-2015 with less than 60% cloud cover and downloaded from US Geological Survey (USGS), European Space Agency (ESA) and the Google Cloud Storage. All cloud/cloud-shadow masks have been obtained from USGS via the ESPA (Earth Resources Observation and Science (EROS) Center Science Processing Architecture) on demand interface which employs a C version of the FMask algorithm. The resulting dataset, for which more than 1.5PB of intermediate products were generated, is referred to as Landsat TimeScan 2015[34]. Specifically, the whole processing has been carried out at the IT4Innovations Czech supercomputing center (Ostrava) in the framework of ESA's Urban Thematic Exploitation Platform (U-TEP)[35] project. Classification has been also carried out in the same infrastructure, whereas post-classification activities have been performed in the Calvalus system[36] available at DLR's Earth Observation Center.

To effectively handle the huge amount of data to process, working units of 1x1 degree size have been defined and the final WSF2015 is obtained as a mosaic of ~14K tiles (where at least a single settlement has been detected).

In Figure 3, an overview of the WSF2015 is given for the entire World, along with 4 different zooms referring to: A) Eastern China and Korea; B) Western Europe; C) Mid-Atlantic USA; and D) the Nairobi region in Kenya. Zoom A refers to one of the most populated regions of the world, including - among others - the Bohai Economic Rim (at the top), as well as the Yangtze River Delta and Pearl River Delta megalopolis (at the center and bottom, respectively). However, yet at this scale it is immediately evident how, besides the major cities, the WSF2015 also outlines the thousands of medium and small-size settlements scattered throughout the whole region, especially in the North China Plain which exhibits an extremely flat topography. This is also evident in Zoom B, where the myriad of towns and (especially) small villages characterizing the Western European landscape are properly mapped. Moreover, one can also start noticing the high detail in the delineation of the bigger cities (e.g., London at the left, Paris at the bottom left, Berlin at the top right), which can be further appreciated in Zoom C. Here, a portion of the US Northeast megalopolis is shown stretching from Washington-Baltimore (bottom left corner), to Philadelphia (center) and to New York (top right corner). The WSF2015 reliably outlines all major centers, as well as their fragmented metropolitan and suburban areas; concurrently, it also detects the small rural villages located in the nooks of the Appalachian Mountains (top left corner). Finally, in Zoom D the layer proves capable of capturing the very complex settlement pattern north of Nairobi (located at the bottom center) which includes the counties of Muranga and Kiambo. Specifically, these are mostly characterized by a rugged landscape interspersed with several hillocks where residents intensively settled along the many valleys in the region (thus resulting in the striped linear pattern that might be falsely interpreted as misclassification at first sight).

Overall, the WSF2015 estimates a global settlement surface of ~1.28 MKm², which corresponds to ~0.95% of the emerged surfaces (i.e., ~134.77Mkm² excluding Antarctica). The dataset has been recently used by the Authors to perform a thorough analysis of the worldwide settlement spatial variability and structure through advanced scaling analysis[37]. Settlement density proved not suitable for explaining alone the high variability of existing patterns, hence a novel global categorization is proposed.

**DATA RECORDS**

The WSF2015 layer described in this article is publicly and freely available through Figshare. The dataset is organized for download in 306 GeoTIFF files (EPSG4326 projection, deflate compression) each one referring to a portion of 10x10 degree size (~1110x1110km) whose upper-left and lower-right corner coordinates are specified in the file name [e.g., the tile WSF2015_v1_EPSG4326_e010_n60_e020_n50.tif covers the area between (10E;60N) and (20E;50N)]. A virtual mosaic file (i.e., WSF2015_v1_EPSG4326.vrt) is also provided which allows visualizing the global product at once on most diffused GIS platforms (e.g., ArcGIS, QGIS, MapInfo). Settlements are associated with value 255; all other pixels are associated with value 0.

Additionally, 5 resampled versions are also provided at 100m, 250m, 500m, 1km and 10km,

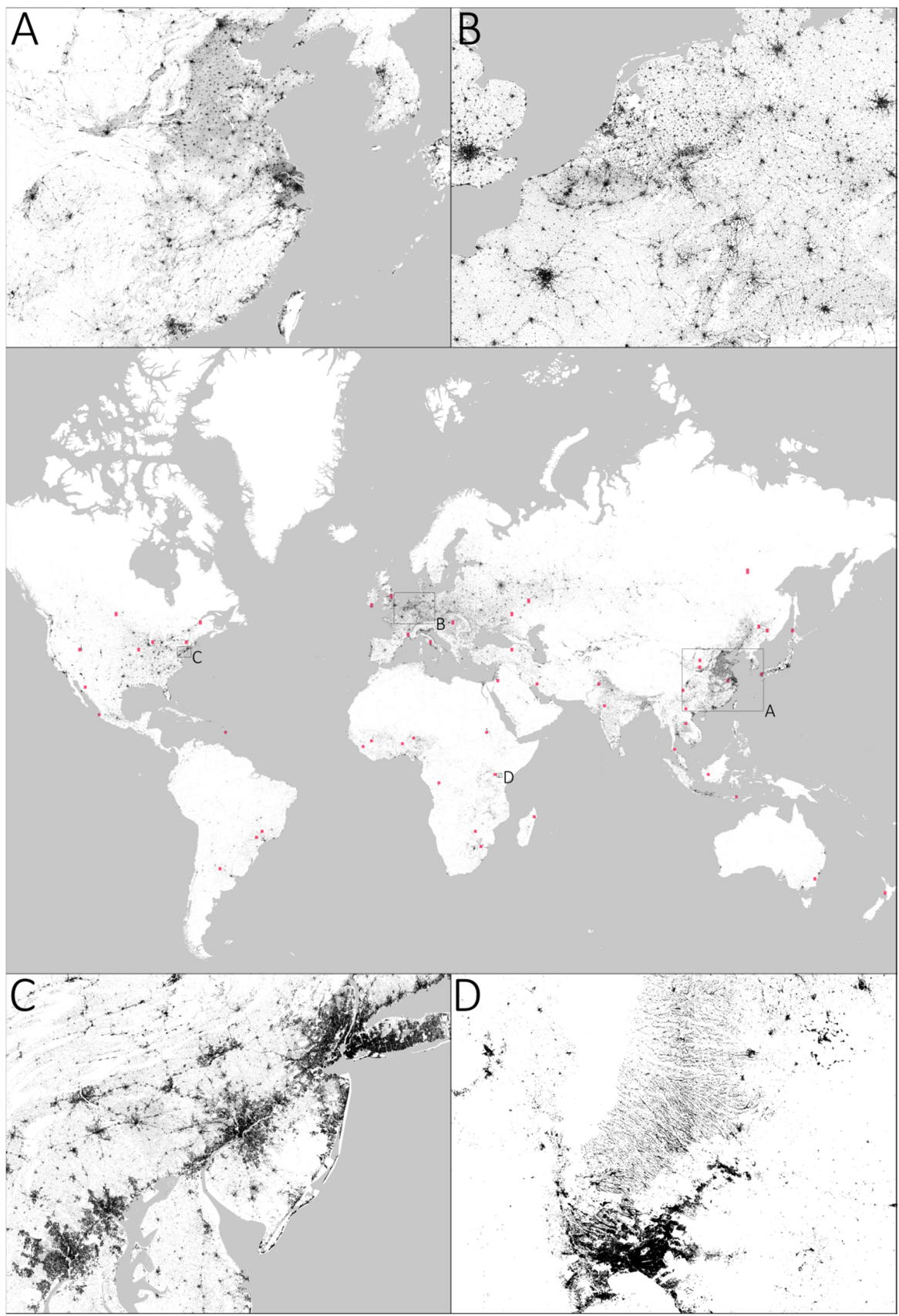

Figure 3. **WSF2015**. Overview of the WSF2015 for the entire World, along with 4 different zooms referring to: A) Eastern China and Korea; B) Western Europe); C) Mid-Atlantic USA; and D) the Nairobi region in Kenya. Validation sites selected for assessing the quality of the layer are reported as red squares.

respectively, reporting for each pixel the corresponding ground percent surface covered by settlements. These can be efficiently used, for instance, as input to regional, continental, or global models and are distributed as individual GeoTIFF files embedding overviews for the levels 2, 4, 8, 16, 32, 64, 128 and 256.

**TECHNICAL VALIDATION**

In the framework of remote sensing, accuracy assessment is generally separated into three major components[38], namely:

- response design, which defines the protocol for determining whether the map and reference classifications are in agreement;
- sampling design, which defines the protocol for identifying a representative subset of the region under analysis (given the impossibility of applying the response design to the entire classification map);
- analysis, which defines how to quantify accuracy.

In the following, the strategy designed for validating the WSF2015 is presented; in particular, specific details are given about the protocols adopted for each of the abovementioned components, while final results are discussed afterwards.

**Response Design**

The four major features of the response design include the source of information from which reference data are taken, the spatial unit, the labeling protocol for the reference classification, and a definition of agreement:

- ***Source of Reference Data:*** Google Earth satellite/aerial VHR imagery available for the period 2014-2015 has been used. The spatial resolution varies depending on the specific data source; in the case of SPOT imagery it is ~1.5m, for Digital Globe's WorldView-1/2 series, GeoEye-1, and Airbus' Pleiades it is in the order of ~0.5m resolution, whereas for airborne data (mostly available for North America, Europe and Japan) it is about 0.15m.
- ***Spatial Assessment Unit:*** since input data with different spatial resolutions have been employed to generate the WSF2015 (i.e., 30m Landsat-8 and 10m S1), a 3x3 block spatial assessment unit composed of 9 cells of 10x10m has been chosen.
- ***Reference Labeling Protocol:*** in our study we define:
    - *building* as any structure having a roof supported by columns or walls and intended for the shelter, housing, or enclosure of any individual, animal, process, equipment, goods, or materials of any kind;
    - *building lot* as the area contained within an enclosure (e.g., wall, fence, hedge) surrounding a building or a group of buildings;
    - *road* as any long, narrow stretch with a smoothed or paved surface, made for traveling by motor vehicle, carriage, etc., between two or more points;
    - *paved surface* as any level horizontal surface covered with paving material.
  
  Based on this taxonomy, 4 possible labels have been defined, namely:
    o Buildings: if the given cell intersects any building;
    o Building Lots: if the given cell intersects any building lot and no buildings;
    o Roads/Paved-Surfaces: if the given cell intersects any road/paved surface and no buildings or building lots;
    o None of the previous.
  
  The labelling task has been performed by crowdsourcing internally at Google. Specifically, by means of an a*d-hoc* tool, operators have been iteratively prompted a 3x3 assessment unit on top of the available Google Earth reference VHR scene closest in time to the year 2015 and given the possibility of assigning any of the 4 labels defined above to each cell. For training the operators, a representative set of 100 reference 3x3 units was prepared in collaboration between Google and DLR.

- *Defining Agreement:* to cope with the different existing definitions of settlement, we computed the assessment figures by separately considering as settlement all areas covered by: i) buildings; ii) buildings or building lots; and iii) buildings, building lots or roads/paved-surfaces. Furthermore, 4 different agreement criteria have been defined (see Figure 4), specifically:
    1) for each cell, positive agreement occurs only for matching labels between the classification and the reference;
    2) for each block, a majority rule is applied over the entire 3x3 block of both the classification and the reference; if the final labels match, then the agreement is positive;
    3) for the classification, a majority rule is applied over the entire 3x3 block; for the reference, each block is labelled as settlement only if it contains at least one cell marked as settlement; if the final labels match, then the agreement is positive;
    4) for both the classification and the reference, each block is labelled as settlement only if it contains at least one cell marked as settlement; if the final labels match, then the agreement is positive.

**Sampling Design**

As recommended in the state-of-the-art good practices for assessing land-cover map accuracy[39,40], stratified random sampling design has been chosen. In particular, it is a probability sampling design and one of the easiest to implement; indeed, it involves first the division of the population (i.e., the collection of all pixels contained in the map) into mutually exclusive subsets (i.e., strata) within which random sampling is performed afterwards.

To include a representative set of settlement patterns, 50 tiles of 1x1 degree size (out of the ~14.000 composing the WSF2015) have been selected based on the ratio between the number of settlements (i.e., disjoint clusters of pixels categorized as settlement in the WSF2015) and their overall area. In particular, the *i*-th selected tile has been chosen randomly among those whose ratio belongs to the interval $]P_{2(i-1)}; P_{2i}], i \in [1; 50] \subset \mathbb{N}$ (where $P_x$ denotes the *x*-th percentile of the ratio). The final selected tiles are shown in red in Figure 3.

As the settlement class covers a sensibly smaller area compared to the merger of all other non-settlement classes, an equal allocation reduces the standard error of its class-specific accuracy. Moreover, such an approach allows to best address user's accuracy estimation, which corresponds to the map "reliability" and is indicative of the probability that a pixel classified on the map actually represents the corresponding category on the ground[41,42]. Accordingly, for each of the 50 selected tiles we randomly extracted 1,000 settlement and 1,000 non-settlement samples from the WSF2015 and used these as center cells of the 3x3 block assessment units to be labelled by photointerpretation. Such a strategy resulted in an overall amount of (1,000 + 1,000) x 9 x 50 = 900,000 cells labelled by the crowd. To our knowledge, this outnumbers any other similar exercise presented so far in the literature.

**Analysis**

To finally assess the accuracy of the WSF2015, we considered a series of measures commonly employed in the remote sensing community[39], namely:

- the Kappa coefficient[43,44], which jointly takes into account omission (i.e., underestimation) and commission (i.e., overestimation) errors, as well as the possibility of chance agreement between classification and reference maps. Kappa assumes values between -1 and 1 and a common rule-of-thumb for its interpretation is the following[45]: <0 no agreement; 0 - 0.20 slight; 0.21 - 0.40 fair; 0.41 - 0.60 moderate; 0.61 - 0.80 substantial; 0.81 - 1.0 perfect;

- the percent producer's accuracies $PA_S\%$ and $PA_{NS}\%$ of the settlement and non-settlement class, respectively. Specifically, they denote the portion of assessment units (i.e., cells or blocks) categorized as settlement/non-settlement according to the collected reference information which are correctly categorized as settlement/non-settlement in the classification map. Its complementary measure (100 – $PA\%$) corresponds to the percent omission error;

- the percent user's accuracies $UA_S\%$ and $UA_{NS}\%$ of the settlement and non-settlement class, respectively. Specifically, they denote the proportion of all assessment units (i.e., cells or blocks) categorized as settlement/non-settlement in the classification map which are categorized as settlement/non-settlement also according to the collected reference information. Its complementary measure (100 – $UA\%$) corresponds to the percent commission error;

- the percent average accuracy $AA\%$, which is obtained as the mean between $PA_S\%$ and $PA_{NS}\%$ and represent a balanced measure of correct settlement and non-settlement detection.

**Quality Assessment**

Figure 4 reports the accuracies over the 900,000 collected reference samples computed for the WSF2015 and, concurrently, the GUF, GHSL and GLC30 layers for comparison. In particular, results are given for all combinations (overall 12) of three considered settlement definitions and four assessment criteria. Due to the different spatial resolution of the GUF (12m) and both the GHSL and GLC30 (30m), while assessing their quality, each 10x10m cell of the considered block spatial assessment unit is tagged as settlement only if the intersection with the specific layer is positive.

Noticeably, in all experiments the WSF2015 exhibited the best $AA\%$, with a remarkable average of 86.37 and a mean increase with respect to GUF, GHSL and GLC30 of +6.24, +15.28 and +18.58, respectively. Alongside, it resulted in an average Kappa of 0.6885 with a mean increase of +0.0754 with respect to the GUF and, especially, +0.2338 and +0.2975 with respect to GHSL and GLC30, respectively.

By analyzing the numbers into detail, one can notice a noteworthy increase of the WSF2015 Kappa coefficient for assessment criteria 3 and 4 (0.7646 on average) with respect to criteria 1 and 2 (0.6123 on average). This is due to the fact that 30m resolution Landsat imagery has been employed to generate the product. Hence, even if just a portion of the Landsat pixel on the ground intersects any building, building lot or paved surface, this mostly has a considerable effect in the corresponding spectral signature and the pixel tends to be finally categorized as settlement. This is taken into account by assessment criteria 3 and 4, since the entire 30x30m reference block spatial assessment unit is labelled as settlement even if it contains just one cell marked as settlement.

Assessment criteria 1 and 2 should be then considered more suitable for a fair comparison against the GUF given its 12m spatial resolution. In this case, one can appreciate how the $AA\%$ and Kappa reported for the WSF2015 are in line with those exhibited by the GUF, which has been generated from highly expensive 3m resolution commercial TerraSAR-X/TanDEM-X imagery. Instead, assessment criteria 3 and 4 allow a fair comparison against GHSL and GLC30 as they are both derived from Landsat data. Here, the WSF2015 exhibits notable $AA\%$ and Kappa up to 89.33 and 0.7822, respectively, outperforming both GHSL and GLC30 (with an increase always higher than 17 and 0.32, respectively).

From Figure 4, one can also notice that on average results do not significantly vary across the three considered definitions of settlement; however, a proper analysis allows to better understand which one fits best with the different layers. Concerning the WSF2015, the highest accuracies mostly occur when considering as settlement the combination of buildings and building lots. Only for assessment criteria 1 and 2 Kappa is higher when also roads / paved surfaces are included. Indeed, despite generally

15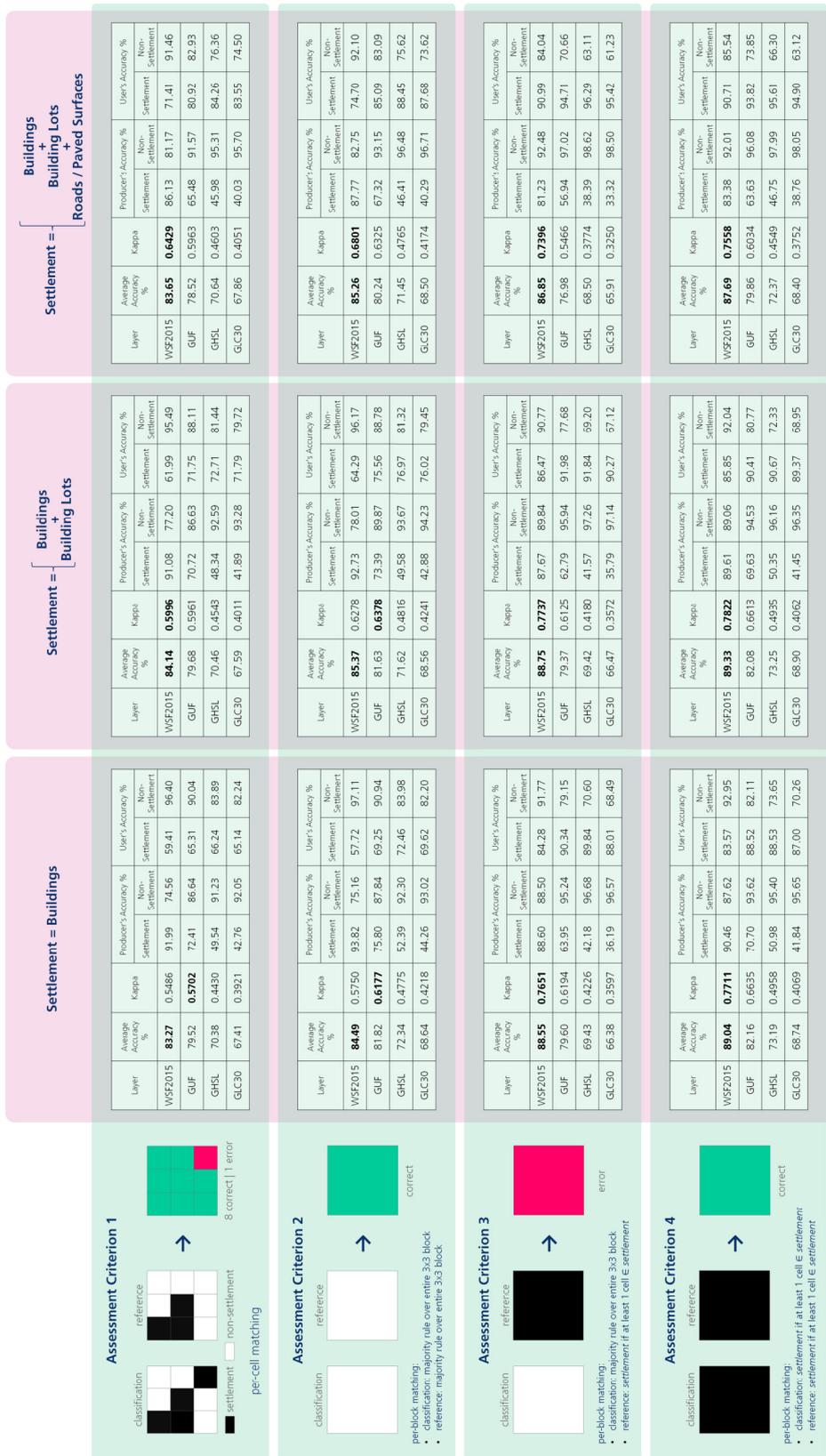

Figure 4. **Quantitative accuracy assessment of the WSF2015 and comparison against the currently most largely employed global settlement extent layers.** Quality assessment figures computed over the 900,000 collected reference samples for the WSF2015, GUF, GHSL and GLC30. Results are concurrently reported for all three settlement definitions and four assessment criteria considered in terms of percent average accuracy, Kappa coefficient, as well as percent producer's and user's accuracies for both the settlement and non-settlement classes.

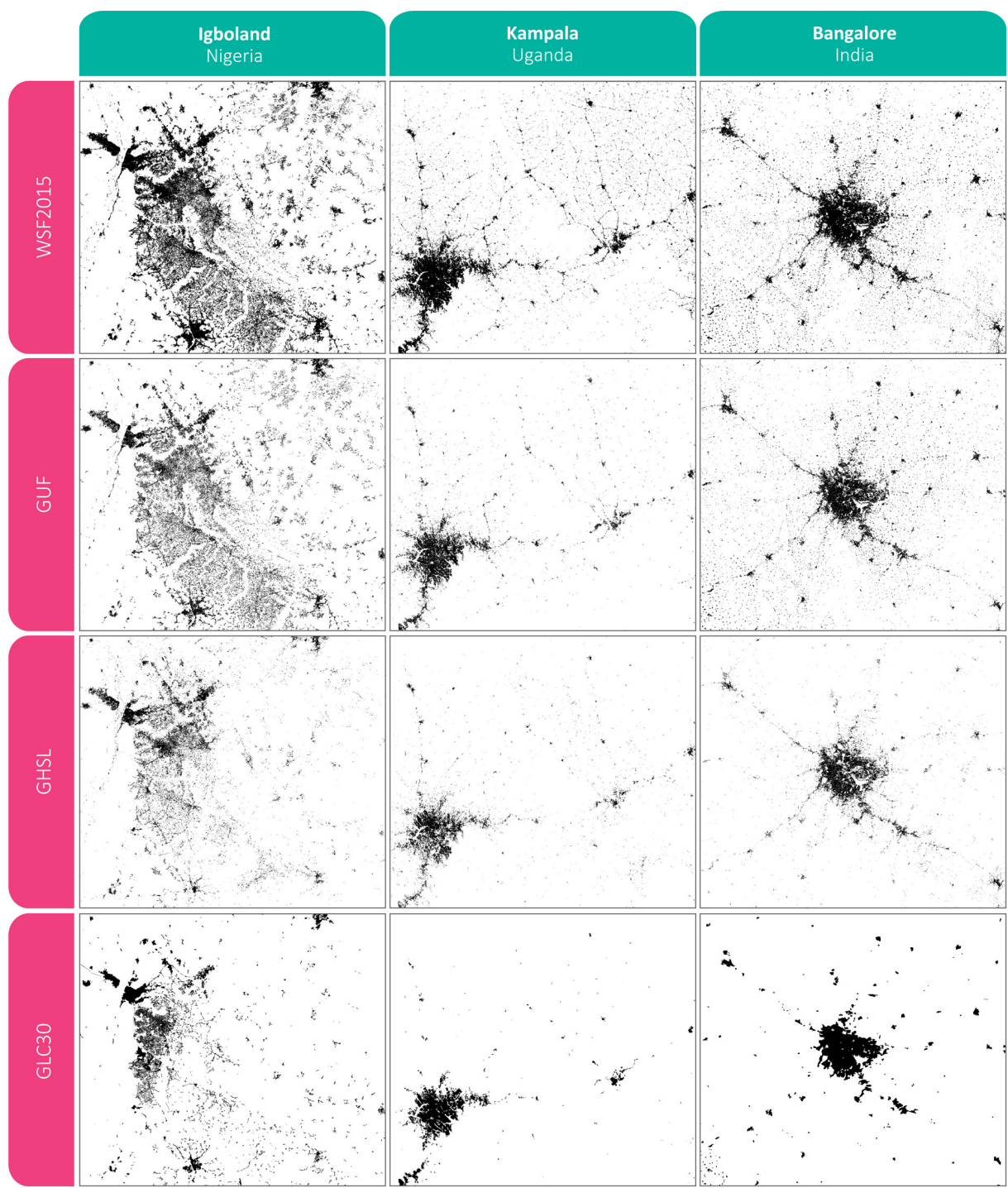

Figure 5. **Qualitative cross comparison of the WSF2015 against the currently most largely employed global settlement extent layers**. Samples for the WSF2015, GUF, GHSL and GLC30 are reported for the Igboland (Nigeria), Kampala (Uganda) and Bangalore (India) regions.

associated with very low S1 backscattering values, most of these are not masked out given their fine scale within urban areas. As regards the GUF, highest *AA%* and Kappa occur partly when only buildings and partly when buildings and building lots are considered as settlement. This is in line with the theory, since the layer has been generated from radar imagery which is sensitive to vertical structures (these comprise both buildings, as well as main elements delimiting building lots like walls, fences, hedges, etc.). In the case of GHSL and GLC30, the two layers show a similar behavior and provide on average a slightly higher Kappa when settlements are defined as combination of buildings and building lots.

Giving a closer look to producer's and user's accuracies it is possible to better understand the nature of the different performances. All GUF, GHSL and GLC30 generally show very high $PA_{NS}\%$ (i.e., >85), but mostly exhibit consistently lower $PA_S\%$, with values never greater than 75.80, 52.39 and 44.26, respectively. On the contrary, the WSF2015 scores overall remarkably high $PA_S\%$ and $PA_{NS}\%$ (on average 88.71 and 84.04, respectively) and, concurrently, it always shows the best $UA_{NS}\%$ in front of a $UA_{NS}\%$ only marginally lower than that of the other layers (on average 92.15 and 75.95, respectively). This quantitatively assesses the capability of the WSF2015 to effectively detect the presence of a considerable number of settlements actually unseen in the other global products. This occurs at the price of a minor settlement extent overestimation mostly due to the employment of the COV texture features; specifically, these allow a more accurate detection in rural and suburban areas, but sometimes result in an overestimation of 1-2 pixels around the actual settlement.

The improved detection performances of the WSF2015 can be qualitatively appreciated in Figure 5, where a cross-comparison against GUF, GHSL and GLC30 is reported for three representative regions including the Igboland (i.e., a cultural and common linguistic region located in south-eastern Nigeria), Kampala (i.e., the capital and largest city of Uganda) and Bangalore (i.e., the capital of the Indian state of Karnataka). Despite the rather different settlement patterns, all three sites are characterized by the presence of medium and large size cities surrounded by a number of very small settlements. As one can notice, the WSF2015 proves extremely effective in all three cases, outperforming all other layers; specifically, it is capable of detecting a higher amount of small villages and better outlining the fringes of major urban areas. The GUF performs equally good only in the Igboland region, but detects considerably less settlements in the Bangalore and, especially, the Kampala case studies. Both GHSL and GL30 exhibit severe underestimation in all three test sites.

### USAGE NOTES

The WSF2015 will be a valuable product in support to all applications requiring detailed and accurate information on human presence. In particular, combined either with other EO or non-EO-based datasets (e.g., related to climate, health, economy, demography, etc.), it will enable deriving indices and metrics of help not only for scientific research but even decision making.

As assessed by the extensive validation exercise, the WSF2015 proved to be the currently most accurate and reliable product of its kind and will hence allow to improve any type of analysis carried out so far with other existing similar layers. Nevertheless, it is worth pointing out that - due to limitations specific of the data used - it was not feasible to consistently detect very small structures (e.g., huts, shacks, tents) because of their reduced scale, the specific building material employed (e.g., cob, mudbricks, sod, straw, fabric), their temporal nature (e.g., nomad or refugee camps), or the presence of dense vegetation preventing their identification.

### CODE AVAILABILITY

The WSF2015 is the result of several processing steps involving tens of sub-modules run on multiple architectures and using different software. While S1 pre-processing and feature extraction has been supported by Google through its Earth Engine platform, the computation of Landsat-8 temporal statistics, the training point extraction and classification tasks have been performed in the IT4Innovations Czech supercomputing center by means of DLR proprietary software, GDAL (Geospatial Data Abstraction Library v.2.4) and Pktools (Processing Kernels for geospatial data v2.6) scripts. Post-classification has been carried out in the Calvalus system available at DLR's Earth Observation Center by means of proprietary software and dedicated Python (v3.5) scripts. Given the use of proprietary tools, the code cannot be openly released to the public.

**ACKNOWLEDGEMENTS**

The Authors wish to acknowledge the European Space Agency (ESA), which supported the implementation and processing of the WSF2015 layer through the SAR4Urban and Urban-Thematic Exploitation Platform (U-TEP) projects, respectively.

**AUTHOR CONTRIBUTIONS**

M.M. coordinated the activity. He designed the whole processing system and validation protocol, as well as supervised and personally supported all the technical processing and quality check tasks. A.M-M. and T.E. provided technical and scientific feedback. U.S. coordinated the data processing with the support of J.Z. A.M-M., D.P.-L., W.H., F.B. and E.S. contributed to the extensive quality check. N.G. supported the S1 data processing. A.K. coordinated the crowdsourcing data collection.

M.M. wrote the manuscript, which has been revised by A.M-M.